\documentclass[12pt]{article}
\usepackage{amsmath, amssymb,graphicx, psfrag, color}

\newwrite\ffile\global\newcount\figno \global\figno=1

\def\writedef#1{}

\input epsf
\def\figin{\epsfcheck\figin}\def\figins{\epsfcheck\figins}
\def\epsfcheck{\ifx\epsfbox\UnDeFiNeD
\message{(NO epsf.tex, FIGURES WILL BE IGNORED)}
\gdef\figin##1{\vskip2in}\gdef\figins##1{\hskip.5in}
\else\message{(FIGURES WILL BE INCLUDED)}%
\gdef\figin##1{##1}\gdef\figins##1{##1}\fi}

\def\figinsert{}
\def\ifig#1#2#3{\xdef#1{fig.~\the\figno}
\writedef{#1\leftbracket fig.\noexpand~\the\figno}%
\figinsert\figin{\centerline{#3}}\medskip\centerline{\vbox{\baselineskip12pt
\advance\hsize by -1truein\center\footnotesize{  Fig.~\the\figno.} #2}}
\bigskip\endinsert\global\advance\figno by1}
\def\endinsert{}

\usepackage{graphics}

\begin{document}

\baselineskip 18pt
\newcommand{\Tr}{\mbox{Tr\,}}
\newcommand{\beq}{\begin{equation}}
\newcommand{\eeq}{\end{equation}}
\newcommand{\bea}{\begin{eqnarray}}
\newcommand{\eea}[1]{\label{#1}\end{eqnarray}}
\renewcommand{\Re}{\mbox{Re}\,}
\renewcommand{\Im}{\mbox{Im}\,}
\newcommand{\yms}{${YM^*\,}$}

\newcommand{\beaa}{\scriptsize\begin{eqnarray}}
\newcommand{\eeaa}{\end{eqnarray}}

\def\N{{\cal N}}


\thispagestyle{empty}
\renewcommand{\thefootnote}{\fnsymbol{footnote}}

{\hfill \parbox{4cm}{
        SHEP-06-26 \\
}}

\bigskip

\begin{center} \noindent \Large \bf
Perfecting the Ultra-violet of Holographic \\ Descriptions of QCD
\end{center}

\bigskip\bigskip\bigskip

\centerline{ \normalsize \bf Nick Evans, Andrew Tedder
\footnote[1]{\noindent \tt
 evans@phys.soton.ac.uk, ajmt@phys.soton.ac.uk} }

\bigskip
\bigskip\bigskip

\centerline{ \it School of Physics and Astronomy} \centerline{ \it
Southampton University} \centerline{\it  Southampton, SO17 1BJ }
\centerline{ \it United Kingdom}
\bigskip

\bigskip\bigskip

\renewcommand{\thefootnote}{\arabic{footnote}}

\centerline{\bf \small Abstract} We investigate imposing a UV
cutoff into a simple AdS/QCD model of the rho mesons. The cutoff
corresponds to the scale at which QCD moves from perturbative to
non-perturbative behaviour, above which the gravity dual will
itself become strongly coupled. Simply imposing a cutoff
significantly improves the fit to the masses of the tower of
excited rho mesons. Formally one should match the couplings of
higher dimension operators and the anomalous dimensions of fields
to the QCD values at the cutoff. We explore examples of these
matchings including looking at the anomalous dimensions of the
$\bar{q} \gamma^\mu q$ operators and including a $G Tr F^4$
coupling.

\newpage


\section{Introduction}

The AdS/CFT Correspondence \cite{Mal,Wit} has provided a new
holographic methodology for computing in strongly coupled gauge
theories. Amongst the properties of QCD-like gauge theories that
have been understood in this arena are confinement
\cite{Maldacena:1998im} quarks \cite{hep-th/0205236}, chiral
symmetry breaking \cite{BEEGK}, and the meson spectrum
\cite{Mateos}. Most recently, phenomenological models of QCD in
this spirit have been constructed
\cite{Erlich:2005qh}-\cite{Csaki:2006ji} which agree with QCD
meson data at better than the 20$\%$ level.

These models which are classical theories of gravitational and
gauge degrees of freedom are all built around asymptotically AdS
spaces which extend out to infinite radius. The radial direction
in the gravity dual corresponds to energy in the gauge theory. In
the rigorous AdS/CFT setting these theories are dual to a
conformal strongly coupled gauge theory (usually the ${\cal N}=4$
super Yang-Mills theory) with relevant operators. Their return to
the strongly coupled fixed point in the UV sustains the weakly
coupled gravity description to infinite radius.

In reality a gravitational dual of QCD would only describe the
strong coupling regime. Where asymptotic freedom drives QCD to
weak coupling, the gravity dual would become strongly coupled
itself. One should therefore impose an ultra-violet, large radius,
cutoff on the gravity dual. To truly describe QCD one should
perform a matching of the gravity dual to QCD in this transition
regime in energy; a regime where strong coupling effects might be
expected to be setting in already in the field theory. On the
gauge side one should match the anomalous dimensions of fields,
the couplings of higher dimension operators and also include all
operator expectation values. As we pointed out in
\cite{Evans:2005ip} if all these matchings could be performed one
would have a description of QCD in the spirit of perfect lattice
actions \cite{hep-lat/9308004}.

This matching seems a hard task, not least because there are
potentially an infinite tower of couplings of higher dimension
operators possible, but the success of the simple AdS models
suggests that deviations from matching to perturbative QCD are not
very large. One can hope to follow the path of improved lattice
actions and try to identify the most significant deviations from
the AdS picture that are needed to better match QCD. One can use
QCD data to tune matching parameters and then hopefully further
predictions will be more accurate. We previously investigated
these ideas \cite{Evans:2005ip} in the glueball sector of a pure
Yang Mills theory but the paucity of data removed any ability to
test the predictions.

Here we will attempt to implement these ideas in the rho meson
sector of QCD. There is a reasonable amount of data provided by
the masses of the excited states in this sector and the AdS/QCD
description \cite{Karch:2006pv} is very simple so there are
relatively few matching parameters that can be tuned. We will show
that just including a UV cutoff makes a substantial improvement to
the fit to the meson masses.

We also study how the theory can be brought closer to ``perfect"
by adjusting the $\bar{q} \gamma^\mu q$ operator's anomalous
dimension. In addition, we consider the inclusion of a coupling of
a higher dimension operator in the glue sector of the theory. In
both cases a fit to the data suggests these are small effects.
Here one is beginning to loose predictiveness again and instead
the results are better interpreted in terms of learning about the
matching conditions from the data.

\section{$\rho$ Mesons in AdS/QCD}

We will use the simple AdS/QCD model of \cite{Karch:2006pv} as our
starting point. This model has the metric and dilaton

\beq ds^2 =  ( \frac{dr^2}{r^2} + r^2\eta_{\mu \nu} dx^\mu dx^\nu),
\hspace{1cm} \Phi(r) = r^{-2} \eeq

The non zero dilaton is present to provide both an IR cut off and
to ensure the tower of $\rho$ meson masses grow like $\sqrt{n}$
with $n$ the excitation number. This behaviour is that expected
from simple confinement models \cite{shif}. The $\rho$ mesons are
described by a vector field in this space

\beq S = \int d^5x \sqrt{g} e^{-\Phi}\textrm{Tr}
\left( - \frac{1}{4 g_5^2} \left(F_L^2+F_R^2\right) \right) \eeq

One seeks normalizable solutions of the linearised equation of motion for
$A^\mu_V $ of the form

\beq  A^\mu_V ~=~ f_n(r) \rho_n^\mu (x) ~=~ A_{0}^\mu f_n(r) e^{i
k.x}, \hspace{1cm} k^2 = -M_n^2 \eeq where $A_V=(A_L+A_R)/2$ and
$A^\mu_0$ is a constant.
\smallskip

 $f_n(r)$ must satisfy

\beq \partial_r (r^3 e^{-\frac{1}{r^2}} \partial_r f_n) + M^2_n
\frac{e^{-\frac{1}{r^2}}}{r} f_n = 0 \label{EoM}\eeq

The normalized solutions have been shown to be

\beq f_n = \frac{1}{r^2} \sqrt{ 2  \over 1+n} L_n^{(1)}(r^{-2}) \eeq where
$L_n^{(m)}$ are the associated Laguerre polynomials. The squared masses are

\beq M_n^2 = 4 (n+1) \eeq

The first five $\rho$ meson excitations predicted by the model
(and scaled to the lightest $\rho$ mass) are

\begin{center}
\begin{tabular}{c|c|c}
$n$ & $m_\rho/$MeV in QCD & $m_\rho/$MeV in AdS/QCD \\
\hline
$\rho$ & 776 & 776 \\
$\rho^*$ & 1459 & 1097\\
$\rho^{**}$ & 1720 & 1344 \\
$\rho^{***}$ & 1900 & 1552 \\
$\rho^{****}$ & 2150 &1735 \end{tabular}
\end{center}

The observed QCD values are also listed for comparison. Although
the $\sqrt{n}$ behaviour is reproduced the masses are consistently
low relative to the data (the RMS error, which for $n$ operators
$O$ is given by $\epsilon_{RMS} = ( \sum_O ( {\delta O \over O
})^2 {1 \over n} )^{1/2}$ is 21\%).

\section{A UV Cutoff}

\subsection{$\rho$ masses}
The first act in regularizing the UV of the theory is to put in a
cutoff at large $r$. This cutoff should correspond to the scale at
which QCD becomes non-perturbative or, in the holographic dual,
the scale below which the classical gravity approximation becomes
trustable. We first try to fix the position of this cutoff
phenomenologically by varying it's position and computing the
$\rho$ meson masses. We look numerically for regular solutions of
eqn (\ref{EoM}) as described above but now using the UV boundary
conditions

\beq  f_n(r) \sim r^{-2}, ~~~~ f'_n(r) \sim -2r^{-3} \eeq at the
cutoff rather than at infinite radius. This choice corresponds to
setting the dimension of the operator $\bar{q} \gamma^\mu q$ to be
three at the UV matching scale where QCD starts to become
non-perturbative.

We plot the tower of masses in figure 1 as a function of the
position of the cutoff $\Lambda$, which we assign both a value of
$r$ and the equivalent energy scale in the gauge theory. The $n=0$
state is in each case normalized to the physical $\rho$ mass. The
dots represent the experimental values.

\begin{figure}[!h]
\begin{center}
  \includegraphics[height=4.5cm,clip=true,keepaspectratio=true]{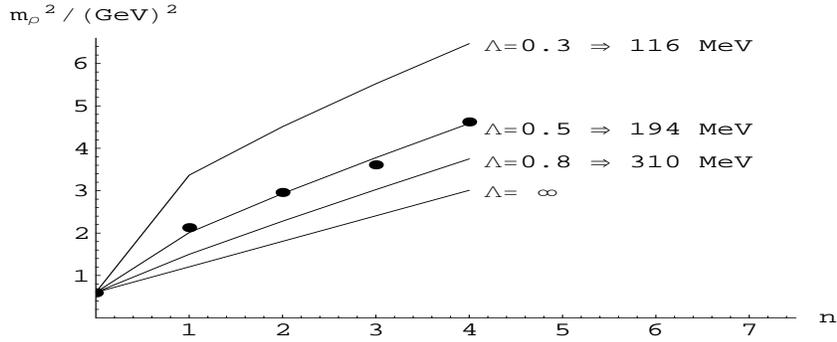}
  \caption{The mass squared of the $n$ lightest $\rho$ meson excitations for
  different values of the UV cut off listed as both a value of
  $r$ in the AdS space and the equivalent energy scale.
  The dots represent the experimental values.}
\end{center}
\end{figure}

Decreasing the cutoff scale from infinite radius raises the masses
of the excited tower states as is required to better match the QCD
data. Significant corrections only set in when $\Lambda$ is small
and the best fit occurs when the cutoff is placed at a radius of
$r = 0.50$. The fit is then extremely good with an RMS error of
1.8\%. The improvement is sufficiently impressive to suggest that
imposing a UV cutoff is the appropriate way to improve AdS/QCD.

The value of the QCD strong coupling scale is set by the
coefficient in front of the $r^{-2}$ term in the dilaton which
explicitly breaks the scale invariance. Throughout we have chosen
the coefficient to be one. Matching our results using the $n=0$
state's mass we find that the best fit value of $\Lambda$
corresponds to a mass scale of just $194$ MeV. Clearly this
suggests that the regime of validity of a gravity dual to QCD is
actually rather small (the same was found in our previous study of
a pure glue theory \cite{Evans:2005ip}). Given the precocious
asymptotic freedom of QCD perhaps this is not so surprising
though. Note the small value of the best fit UV cutoff also means
the gravitational theory has a significant dilaton factor
throughout it's regime of validity.

It is worth pointing out that with this value of $\Lambda$, we
predict the next 3 excited $\rho$ mesons to be at 2320 MeV, 2475
MeV and 2626 MeV. Experimental searches have reached up to 2510
MeV, and so far the highest excited $\rho$ meson found is the
final one listed above at 2150 MeV.

\subsection{Decay constants}

The $\rho$ meson decay constants can be found by substituting the
regular solutions $f_n(r) \rho^\mu(x)$  back into the 5d action
and integrating over $r$. The decay constants are given by
\cite{Erlich:2005qh}

\beq F_{\rho}^2=\frac{1}{g_5^2}\left(\Lambda^3
f_n'(\Lambda)\right)^2 \eeq

Since the large $r$ behaviour of $f_n(r) \sim 1/r^2$ for all $n$,
the different excited states only differ in their decay constants
as a result of the different normalizations of the $f_n$. We
require that the kinetic terms for the different rho excitations
are all canonical which implies imposing

\beq \int_0^{\Lambda} dr  \frac{e^{-r^{-2}}}{rg_5^2}f_n^2 = 1 \eeq

In the original AdS/QCD model with the UV cutoff at infinity one
finds the decay constants grow as the square root of the
excitation number $n$

\beq F_{\rho_n}^2 = { 8 (n + 1) \over g_5^2} \eeq

If one matches $g_5$ to the perturbative high energy vector
correlator \cite{Erlich:2005qh,DaRold:2005zs} so $g_5^2 = 12
\pi^2/ N_c$ then the $n=0$ $\rho$ has a decay constant
$F_{\rho_0}^{1/2} = 260$ MeV compared to the physical value of
$345$ MeV.

In figure 2 we display the results of the same computation with a
UV cutoff present. For low cutoffs $F_{\rho_0}$ rises: with
$\Lambda = 194$ MeV $F_{\rho_0}^{1/2}=478$ MeV. Comparison with
the physical value again hints that a low cutoff is appropriate.
On the other hand, as the cutoff is brought down the $\sqrt{n}$
behaviour (argued for in \cite{shif}) is apparently lost and the
higher resonance decay constants fall relative to the $n=0$ case.
The reason for this is that the cutoff impedes on the values of
$r$ where the wave functions of the eigenstates are substantial.
By the time that the cutoff is of order a few hundred MeV the
integral for the normalization is dominated around the cutoff.
This makes the computation of the decay constant suspect -
formally one needs a description of the physics to higher energies
which may lie beyond the region of perturbative validity for the
supergravity. Note this contrasts with the computation of the
masses - those values are determined by requiring regular
solutions in the infra-red away from the cutoff.

\begin{figure}[!h]
\begin{center}
  \includegraphics[height=4.5cm,clip=true,keepaspectratio=true]{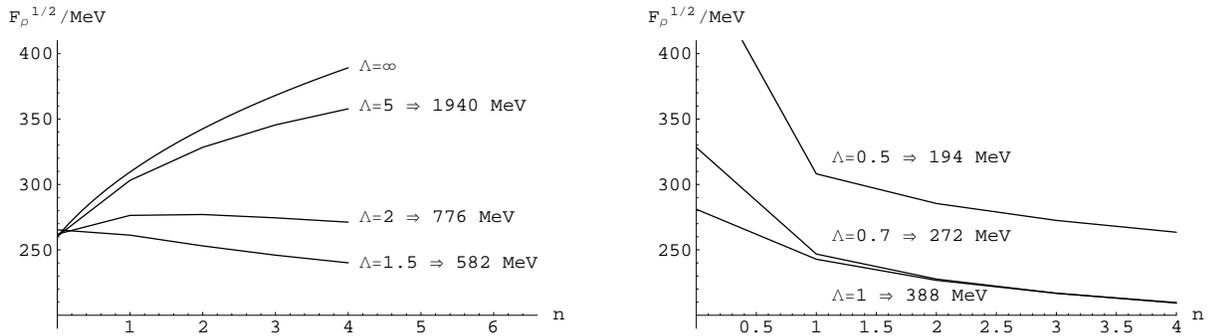}
  \caption{The decay constant $F_\rho^{1/2}$ for the different $\rho$ excitations
  plotted for varying UV cut off.}
\end{center}
\end{figure}

\section{Anomalous Dimensions at the Cutoff Scale}

If one is going to impose a UV cutoff on the gravitational
description that lies in the transition region where QCD moves
from perturbative to non-perturbative behaviour, one should be
careful in matching the theories. It is possible that, for
example, substantial non-perturbative effects should be included:
in other words, the dimension of the operator $\bar{q} \gamma^\mu
q$ should not be simply set to three. In the gravity dual this
dimension is encoded by the initial conditions set for the
eigenfunctions at the cutoff.

One can repeat the computation of the $\rho$ masses using boundary
conditions

\beq  f_n(r) \sim r^{-\omega}, ~~~~ f'_n(r) \sim - \omega
r^{-\omega-1}\eeq at $r=\Lambda$ with $\omega$ a free parameter.

We find that for all values of the cutoff, from $\infty$ down to
our best fit value, the preferred value of $\omega$ is two, which
is consistent with the naive dimension matching. This appears to
tell us that the anomalous dimensions are not large.

However, if one evaluates the derivative of the flows $f_n(r)$ at
a finite values of $r$ in the case of the model with a UV cut off
at infinity then the derivative differs for each value of $n$.
This suggests that formally one must allow $\omega$ to vary state
by state. Of course this introduces so many free parameters that
predictions are lost. It is nevertheless interesting to see how
much these anomalous dimensions must be shifted from two to
precisely reproduce the observed spectrum. For example for our
best fit value of $\Lambda=194$ MeV one finds

\begin{center}
\begin{tabular}{c|c}
meson  & $\omega$ \\ \hline
$\rho^*$ & 1.85 \\
$\rho^{**}$ & 1.98 \\
$\rho^{***}$ & 2.12 \\
$\rho^{****}$ & 1.98  \end{tabular}
\end{center}

\noindent The corrections are less than $10\%$.

\section{Coupling of a Higher Dimension Operator}

The presence of large couplings for higher dimension operators
before QCD can be matched to the perturbative gravity theory would
be another signal of non-perturbative phenomena. There are
formally an infinite set of such couplings which might be
important. One might hope the couplings of lower dimension
operators would grow fastest as one moved into the
non-perturabtive regime though. Such couplings, which are
irrelevant perturbations, will appear in the gravitational dual as
deformations of the metric which grow at large $r$. We know how to
encode one simply \cite{Evans:2005ip,hep-th/0112058} so will
investigate the effect of that.

If we write the metric as

\beq ds^2 =  H^{1/2}dr^2 + H^{-1/2} \eta_{\mu \nu} dx^\mu dx^\nu
\eeq Then we can deform the AdS space by allowing it to return to
flat space asymptotically,

\beq H(r) \rightarrow r^{-4} + \alpha = r^{-4}\left(1+\alpha r^{4}\right) \eeq

The parameter $\alpha$ is a symmetry singlet and has energy
dimension -4. It should therefore be identified with the coupling
of the term $G Tr F^4$.

We have repeated the fit to the lightest five $\rho$ meson masses
in this deformed geometry. The preferred value of $\alpha$ as a
function of $\Lambda$ and the RMS error of the fit are

\begin{center}
\begin{tabular}{cc|c|c}
$\Lambda $ & & $\alpha$ & $\epsilon_{rms}$  \\ \hline
$r=20$  & (1137 MeV) & 0.0011  & 8.18 \%\\
$r=10$ & (5606 MeV) & 0.0045 & 7.51 \% \\
$r=5$ &  (2718 MeV) & 0.019  & 6.77 \%\\
$r=1$ &  (469 MeV) & 0.75  & 4.70 \% \\
$r=0.5$ & (194 MeV)& 0.08 & 1.83\% \end{tabular}
\end{center}

Note $\alpha$ grows as $\Lambda$ is lowered - this is natural
since $\alpha$ changes the large $r$ part of the metric. To change
the results when only a small $r$ slice of the metric is
considered needs a large $\alpha$.

It can be seen that the fit at a given value of $\Lambda$ is
improved by the inclusion of $\alpha$. Amusingly as one approaches
the best fit value of $\Lambda$ we found above, the preferred
value of $\alpha$ suddenly becomes very small. This is a
reflection of just how good the fit is from just including a
cutoff. We conclude that if an appropriately low cutoff is
included the $G Tr F^4$ coupling is in fact a small effect.

\section{Discussion}

A perturbative gravitational dual of QCD should only be expected
to work at energies below a few GeV at best, where QCD is
non-perturbative. We have investigated imposing a UV cutoff on an
AdS/QCD model of the $\rho$ mesons and found that the data has a fit
at the $2\%$ level with a UV cutoff of a few hundred MeV
(compared to a fit of $21\%$ with an infinite cutoff). We
conclude that the holographic description of QCD should only be
used at low energies on a quite small radial interval.

We have also looked at fitting corrections to the anomalous
dimension of the operator $\bar{q} \gamma^\mu q$ and introducing a
coupling of the operator $Tr F^4$. Although these corrections
could be used to fine tune the fit by a percent or so they do not
appear to be significant corrections to the model. Of course these
are only easily implementable examples from an infinite set of
possible corrections but finding the corrections to be small
provides further understanding of the success of the basic AdS/QCD
models. One could also try to include the vacuum expectation
values of more operators in the metric (see for example
\cite{Csaki:2006ji}) and a dynamical, predictive mechanism of
chiral symmetry breaking \cite{Evans:2006dj}. Such effects would
be important to study the pion and axial vector meson sectors of
the model. As explained in \cite{Karch:2006pv} the model used here
does not give a good prediction of these sectors because the
dilaton form, put in to give the $\sqrt{n}$ rise in masses, does
not lead to a sensible condensate prediction. If one attempted to
tackle all of these problems then most likely the number of free
parameters would rise faster than the number of available data
points. Of course this reflects the fact that a perfect action is
in the end just a reparametrization of the full QCD spectrum. We
hope though that we have identified the imposition of a UV barrier
as an important correction and that these other effects are
subleading in the $\rho$ sector. Putting together a complete model
of all sectors including the baryons remains as an important
challenge.

\end{document}